# ICU Delirium Prediction Models: A Systematic Review

Matthew M Ruppert, BS[1,4]; Jessica Lipori, BA[1,4]; Sandip Patel, MD[1,4]; Elizabeth Ingersent[1,4]; Tezcan Ozrazgat-Baslanti, PhD[1,4]; Tyler Loftus, MD[3,4]; Parisa Rashidi, PhD[2,4]; Azra Bihorac, MD MS[1,4]

[1]Department of Medicine, University of Florida, Gainesville, FL, USA
[2]Department of Biomedical Engineering, University of Florida, Gainesville, FL, USA
[3]Department of Surgery, University of Florida, Gainesville, FL, USA
[4]Precision and Intelligent Systems in Medicine (Prisma[P]), University of Florida, Gainesville, FL, USA

**Correspondence can be sent to:** Azra Bihorac, MD, MS, FASN, FCCM, Department of Medicine, Division of Nephrology, Hypertension, & Renal Transplantation, 1600 SW Archer Road, PO Box 100224, Communicore Building, Room CG-98, Gainesville, FL 32610-0224. Office Phone (352) 273-9009; Cell Phone: (352) 870-0090; Fax: (352) 392-5465; E-mail: abihorac@ufl.edu

**Reprints:** Reprints will not be available from the author(s).

**Conflicts of Interest and Source of Funding:** This work is supported by NIH/NIGMS RO1 GM-110240 (MR, TOB, PR, AB), Davis Foundation – University of Florida (MR), NIH/NIGMS P50 GM111152 (TOB, AB), NIH/NIGMS T32 GM-008721 (TL), NSF CAREER 1750192 (PR), and NIH/NIBIB 1R21EB027344-01 (PR, AB). The authors declare that they have no competing interests.

**Ethics Approval and Consent to Participate:** N/A

**Consent for Publication:** N/A

**Availability of Data and Material:** N/A

**Abstract**

**Purpose:** Summarize ICU delirium prediction models published within the past five years.

**Methods:** Electronic searches were conducted in April 2019 using PubMed, Embase, Cochrane Central, Web of Science, and CINAHL to identify peer reviewed studies published in English during the past five years that specifically addressed the development, validation, or recalibration of delirium prediction models in adult ICU populations. Data were extracted using CHARMS checklist elements for systematic reviews of prediction studies, including the following characteristics: study design, participant descriptions and recruitment methods, predicted outcomes, *a priori* candidate predictors, sample size, model development, model performance, study results, interpretation of those results, and whether the study included missing data.

**Results:** Twenty studies featuring 26 distinct prediction models were included. Model performance varied greatly, as assessed by AUROC (0.68-0.94), specificity (56.5%-92.5%), and sensitivity (59%-90.9%). Most models used data collected from a single time point or window to predict the occurrence of delirium at any point during hospital or ICU admission, and lacked mechanisms for providing pragmatic, actionable predictions to clinicians.

**Conclusions:** Although most ICU delirium prediction models have relatively good performance, they have limited applicability to clinical practice. Most models were static, making predictions based on data collected at a single time-point, failing to account for fluctuating conditions during ICU admission. Further research is needed to create clinically relevant dynamic delirium prediction models that can adapt to changes in individual patient physiology over time and deliver actionable predictions to clinicians.

**Take-home message:** Most ICU delirium prediction models developed within the past five years predict delirium occurrence within the entire course of ICU admission using input data collected within 24 hours of ICU admission, which is inconsistent with ICU delirium pathophysiology. Further research is needed to create clinically relevant dynamic delirium prediction models that adapt over time and deliver pragmatic, actionable predictions to clinicians.

## 1. Introduction

Delirium is a transient condition consisting of altered attention and consciousness that is common in hospital settings [1]. Delirium has a particularly high incidence in the ICU, ranging from 25% to 87% [2-4]. Factors associated with increased risk for ICU delirium include, but are not limited to: older age, lower levels of education, history of hypertension, alcohol abuse, higher Acute Physiology, Age, Chronic Health Evaluation (APACHE) II scores, and use of sedative and analgesic medications [3, 5, 6]. The use of benzodiazepines for mechanical ventilation carries a particularly high risk for delirium compared with other sedatives [7, 8]. Environmental factors, including isolation, use of physical restraints, and prolonged exposure to light and sound have also been associated with delirium [9, 10].

ICU delirium is strongly associated with adverse outcomes, including increased hospital length of stay, greater morbidity and mortality, poor cognitive recovery, slower rates of overall recovery, and increased cost of care [2, 11, 12]. This makes prevention and early identification of delirium essential. Delirium assessments such as the Confusion Assessment Method for the Intensive Care Unit (CAM-ICU) and the Intensive Care Delirium Screening Checklist (ICDSC) have been shown to be effective in diagnosing delirium [13, 14] and their use is recommended under current clinical practice guidelines [15]. However, these assessments are sometimes not trusted or understood by ICU staff and are therefore inconsistently applied [16-18].

The use of prediction models has shown promise in predicting several types of delirium, including post-operative and subsyndromal delirium as well as delirium in the ICU. These predictions can be used by clinicians as decision support for preventing and treating delirium [19, 20]. However, clinical adoption of delirium prediction models has been limited, perhaps because model outputs are often not pragmatic or actionable for clinicians. Machine learning techniques may abrogate these weaknesses, but contemporary descriptions of these techniques are sparse. The purpose of this systematic review was to describe and compare ICU delirium prediction models developed within the past five years, summarizing model efficacy and identifying model characteristics that impact clinical applicability.

## 2. Materials and Methods

**2.1 Search Strategy and Selection Criteria**

PubMed, Embase, Cochrane Central, Web of Science, and CINAHL were systematically searched for articles relating to delirium prediction models among adult ICU patients.

An ICU delirium prediction model was defined as any model or algorithm that incorporated at least one clinical factor measured during hospital admission to assign an estimated risk of developing delirium during the ICU stay. Studies that specifically addressed the development, validation, or recalibration of prediction models in adult ICU populations were included. Models that were designed to predict delirium in the context of substance abuse or withdrawal were excluded. Abstract only studies were excluded.

Search terms were tailored to utilize Medical Subject Headings (MeSH) or subject headings embedded in each database. Each search query was the union of three search components: delirium, ICU, and prediction. The first component, delirium, contained delirium associated terms and subject headings with words including but not limited to "delirium", "ICU syndrome", "acute confusion", and "Confusion Assessment Method". The second component, ICU, contained ICU associated terms and subject headings with words including but not limited to "ICU", "Intensive Care Unit", "Critical Care", and "Critically Ill". The third component, prediction, contained prediction associated terms and subject headings with words including but not limited to "predict", "model", "risk", and "risk assessment". A full list of the search terms for each database is available in Supplement A.

In addition to the three query components above, search results were restricted to papers published in English within the past ten years. Database searches were performed on April 25th, 2019. There were 7,296 articles remaining after the search results were compiled and the duplicates removed (Figure 1). To further reduce the number of potential articles, articles published prior to January 1st, 2014 were removed, reducing the number of potential articles to 4,940. The 4,940 articles were divided amongst four authors. Each article's title and abstract were reviewed by two authors independently. Disagreements between authors during title/abstract review were settled by the lead author. Title and abstract review based on the above selection criteria reduced the number of articles to 21.

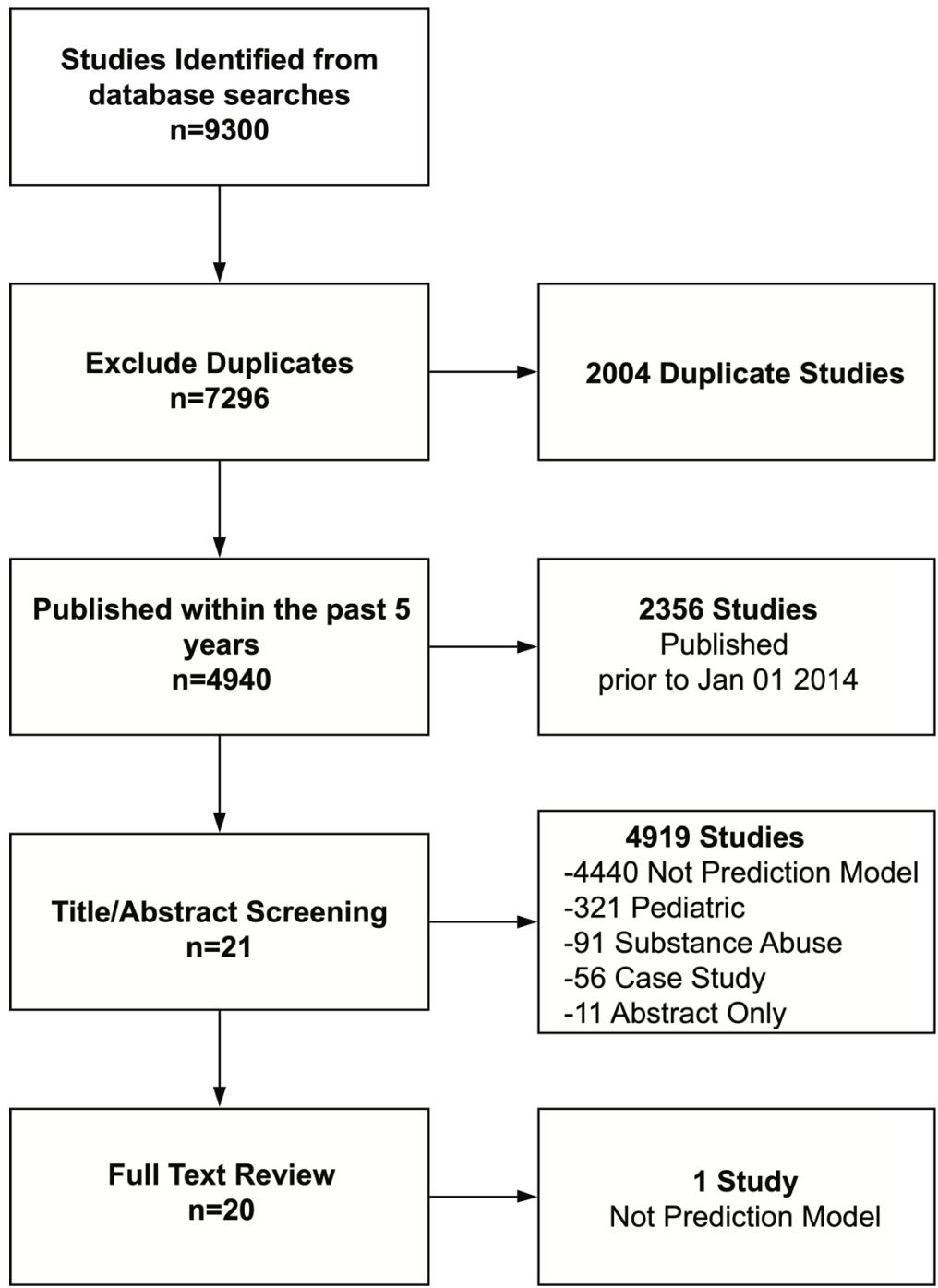

Figure 1: Consort diagram of studies included in review

## 2.2 Data Extraction and Synthesis

The 21 articles were divided into three groups of ten, with a three-article overlap between each group, to verify consistency across authors (42% validation). The data from each article group was then extracted independently by three authors using the CHARMS Checklist for Systematic Reviews of Prediction Studies, excluding the treatments received element [21]. Extracted data elements included study design, participant descriptions and recruitment methods, predicted outcomes, candidate predictors, final predictors, sample size, model development, model performance, model evaluation, study results, interpretation of those results, and treatment of missing data. During data extraction, one article was removed because it was not a true prediction model.

Each model was summarized in terms of: study type (development, validation, or review); study design (prospective or retrospective); cohort division (temporal or random); cohort size; outcome (definition and assessor(s)); risk of bias (outcome assessment method, predictor selection method, handling of missing data, and ratio of delirium events to number of model predictors); *a priori* predictors; predictor selection technique; final model predictors; performance (area under the receiver operating curve (AUROC), accuracy, positive predictive value (PPV), and negative predictive value (NPV)); and model tuning technique.

## 3. Results

### 3.1 Study Characteristics

Of the 20 included studies, 13 were primarily concerned with the development of new prediction models [22-34], six with the validation of existing models [20, 35-39], and one was a review article that summarized delirium prediction models published prior to January 1$^{st}$, 2013 [40] (Table 1). These studies included 26 risk prediction models, of which 18 were developed in the included studies. Of the 13 model development studies, nine were prospective cohorts and four were retrospective cohorts. Sample sizes ranged from 94 to 18,223 participants [28, 34]. Those studies, which validated existing models, had a sample size ranging from 32 to 2,187 participants [37, 38]. Delirium was most commonly assessed using CAM-ICU, though several studies used CAM [25], NuDESC [34], DSM [28], and ICDSC [29].

Table 1. Studies Overview

| Study | Study Type | AUROC |
|---|---|---|
| **Development Studies** | | |
| Chaiwat, 2019 | Prospective | Development: 0.84 |
| | | Internal: 0.82 |
| Chen, 2017 | Prospective | Internal: 0.78 |
| | | External: 0.77 |
| Fan, 2019 | Prospective | Development: 0.89 |
| | | Internal: 0.90 |
| Kim, 2016 | Prospective | Development: 0.911 |
| Marra, 2018 | Prospective | NR |
| Moon, 2018 | Prospective | Internal: 0.72, 0.93, 0.85 |
| Oh, 2018 | Prospective | NR |
| Sakaguchi, 2018 | Retrospective | Development: 0.89 |
| Stukenberg, 2016 | Retrospective | NR |
| Van den Boogaard, 2014 | Prospective | Development: 0.76 |
| | | External: 0.75, 0.713 |
| Wang, 2018 | Prospective | NR |
| Wassenaar, 2015 | Prospective | Development: 0.76 |
| | | Internal: 0.75 |
| | | External: 0.79, 0.68 |
| Wong, 2018 | Retrospective | Development: 0.81 |
| **Validation Studies** | | |
| Azuma, 2019 | Retrospective | PRE-DELIRIC: 0.83 |
| Green, 2019 | Retrospective | PRE-DELIRIC: 0.79 |
| | | E-PRE-DELIRIC: 0.79 |
| | | Chen et al: 0.77 |
| Lee, 2017[a] | Prospective | PRE-DELIRIC: 0.75 |
| | | rPRE-DELIRIC: 0.75 |
| Linakaite, 2018 | Prospective | rPRE-DELIRIC: 0.713 |
| Paton, 2016 | Prospective | NR |
| Wassenaar, 2018 | Prospective | PRE-DELIRIC: 0.74 |
| | | EPRE-DELIRIC: 0.68 |
| **Literature Reviews** | | |
| van Meenen, 2014 | | Bohner et al: NR |
| | | Katznelson et al: 0.76 |
| | | Katznelson et al: 0.77 |
| | | Marcantonio et a: 0.81 |
| | | Kalisvaart- Inouye et al: 0.73 |
| | | Koster et al: 0.75 |
| | | Rudolph et al: 0.74 |

Abbreviations Used: AUROC (Area Under the Receiver Operating Curve), NR (Not Recorded), PRE-DELIRIC (The Prediction Model for Delirium), E-PRE-DELIRIC (The Early Prediction Model for Delirium), rPRE-DELIRIC (The Recalibrated Early Prediction Model for Delirium).

[a] This paper was primarily a validation study, but it also developed a recalibrated version of the model proposed by Katzenlson, 2009 [40].

**3.2 Risk of Bias Assessment**

The major risks of bias in included papers were assessment of outcome, selection of candidate predictors, sample size, and treatment of missing data [21, 41]. Seven studies were at risk for bias due to having multiple delirium assessment measures [28, 34, 36]. Retrospective studies had an increased risk of bias due to both the problem of assessing for delirium retrospectively and that the outcome was often assessed by the same researchers that selected candidate predictors. Selection of predictors was a further source of bias. While some selected candidate predictors from literature review [25-27, 33], many gave little or no justification for candidate predictor selection [22, 24, 29], and some selected final predictors without prior analysis [23]. Most studies did not report missing data or management of missing data. Ten of 18 models that were developed in the included studies were at high risk for overfitting due to a low ratio of delirium incidence compared with the number of candidate predictors [22, 24, 25, 27, 29, 34].

**3.3 Risk Factors**

Six studies evaluated existing delirium prediction models, including the prediction model for delirium (PRE-DELIRIC) and the early prediction model for delirium (E-PRE-DELIRIC). The PRE-DELIRIC model consists of nine risk factors: age, APACHE II score, coma, sedative use, morphine use, serum urea, metabolic acidosis, urgent admission, and admission category (Table S1). The E-PRE-DELIRIC model considered 18 candidate predictors chosen by literature review and input from an expert panel of physicians, and nine of these predictors were included in the final model [34]. The PRE-DELIRIC and E-PRE-DELIRIC share three predictors in common: age, admission category, and urgent admission. Additionally, both utilize some marker of renal function, namely blood urea nitrogen (BUN) and urea concentration. Among other differences, E-PRE-DELIRIC includes a greater number of predictors relating to patient's predisposing factors, such as history of cognitive impairment and history of alcohol use.

Although rationale for candidate predictors was not consistently available, many studies that sought to develop new models for delirium prediction employed a combination of literature review and expert opinion. Some of the more frequently selected candidate predictors were also common to the aforementioned PRE-DELIRIC and E-PRE-DELIRIC models (Figure 2, Table S2). For example, age was included in six studies [22, 23, 25, 26, 31, 33],

APACHE II score was included in four studies [23, 24, 26, 31], and some marker of renal function (blood urea concentration, serum creatinine, or a BUN/Cr ratio) was utilized in three studies [25, 29, 31]. Other commonly considered predictors included mechanical ventilation, urgent admission, and use of antipsychotics, sedatives, and benzodiazepines.

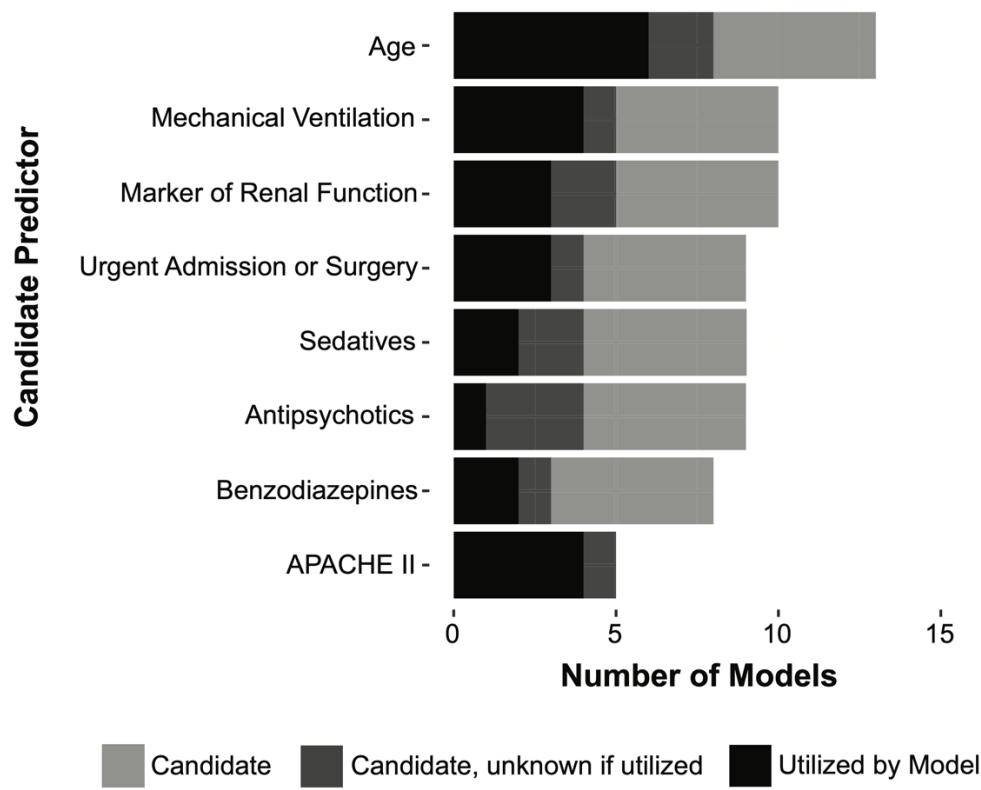

Figure 2: Prevalence of predictors considered in at least five models

Three studies evaluated greater than 40 candidate predictors. Two of these studies employed machine learning to develop delirium prediction models [27, 34]. Another used an informative value calculation and multiple regression analyses "based on literature review" to narrow down the number of predictors in a stepwise manner to use in modeling [25]. A few studies went beyond the existing approach of selecting candidate predictors within 24 hours of admission and instead included various predictors to be collected daily. This was performed in an effort to predict delirium dynamically, and evaluate recurrent or ongoing delirium, which are not assessed by PRE-DELIRIC and E-PRE-DELIRIC.

**3.4 Predictive Model Development and Performance**

Our review included 26 prediction models, of which 18 were developed in the included studies (Tables 2 and S3). These models used various numbers of predictors ranging from one to 588 [32, 34]. Predictors were most commonly chosen for inclusion by logistic regression, though seven models were developed using machine learning techniques [27, 34] and two recalibrated existing models [31, 36]. Of the 17 models developed using logistic regression, six employed additional bootstrapping to allow for better calibration and adjust for overfitting. Multiple methods were employed for determining which variables should be included in multiple logistic regression or the final model. These included pre-selection based on literature review, univariate regression, machine learning techniques, and pre-selection of factors from a previous model. In the final models, most used regression coefficients to establish either a sum score or a score chart with scores stratified into different risk subgroups. Twenty-one of the models measured discrimination with AUROC, reporting values between 0.62 and 0.911 [25, 36]. Studies which statistically assessed the calibration of their models used the Hosmer-Lemeshow goodness of fit test and calibration plots. While most studies were concerned with a binary outcome (delirious vs. non-delirious) occurring at any time, one [26] developed a model aimed at predicting daily transitions between multiple states (normal, delirious, comatose, discharge, or death).

Table 2. Model Performance in Development Studies

| Study (cohort type) | Development/ Validation | Delirium n (%) Development/ Validation | Modelling Approach | Number of Predictors | AUROC (95% CI) Development/Validation |
|---|---|---|---|---|---|
| Chaiwat (P) | 250/ - | 61(24)/ - | MvLR | 5 | 0.84 (0.8- 0.9)/ 0.82 |
| Chen (P) | 310/ 310[a] | 160(26)[b] | MvLR | 11 | -/ 0.78 |
| Fan (P) | 336/ 224[a] | 68(20)/ 46 (20) | mLR | 7 | 0.89 (0.85- 0.93)/ 0.90 |
| Kim (P) | 561/ 553[a] | 112(20)/ 99 (18) | mLR | 9 | 0.911 (0.88- 0.94)/ - |
| Lee (P) [d] | 600/ - | 83(14)/ - | LR | 7 | 0.62/ - |
| Marra (P) | 810/ - | 606 (75)/ - | mLR | 14 | -/- |
| Moon (P) | 3284/ 325[c] -/ 263[a,f] -/ 431[a,f] | 688(21)/ 48 (15) 55(21) 114 (26) | LR[e] | 11 | 0.89/ 0.72 -/0.93 -/0.85 |
| Oh (P) | 94/ - | 39 (41)/ - | LSVM | 1[g] | - /- |
| Sakaguchi (R) | 120/ - | 38(32)/ - | MvLR | 6 | 0.89/ - |
| Stukenberg (P) | 996/ - | 161(16)/ - | LR | 3 | -/- |

| Study (cohort type) | Development/ Validation | Delirium n (%) Development/ Validation | Modelling Approach | Number of Predictors | AUROC (95% CI) Development/Validation |
|---|---|---|---|---|---|
| van den Boogaard (P)[d] | 1824/ - | 363(20)/ - | GLMM | 10 | 0.76 (0.74-0.79)/- |
| Wang (R) | 1692/ - | 32(25) delirium 92(72) delirium and coma | LR | 1 | -/- |
| Wassenaar | 1962/ 952[c] | 480 (25)/ 208 (22) | mLR | 9 | 0.76 (0.73- 0.78)/ 0.75 |
| Wong (R) | 1822/ - | 848(5)/ - | - | - | - /- |
| | | | LR[e] | 114 | Age<65: 0.89; Age>64: 0.80/ - |
| | | | GBM | 345 | Age<65: 0.86; Age>64: 0.80/ - |
| | | | ANN | - | Age<65: 0.71; Age>64: 0.77/ - |
| | | | LSVM | - | Age<65: 0.71; Age>64: 0.76/ - |
| | | | RF | 588 | Age<65: 0.85; Age>64: 0.81/- |
| Bohner[h] | 153 | 60(44) | - | 9 | - |
| Katznelson[h] | 582 | 128(22) | - | 6 | 0.76 |
| Katznelson[h] | 1059 | 122(11) | - | 7 | 0.77 |
| Marcantonio[h] | 1341 | 117(9) | - | 6 | 0.81 |
| Kalisvaart-Inouye[h] | 603 | 74(12) | - | 4 | 0.73 (0.65- 0.78) |
| Koster[h] | 300 | 52(17) | - | 2 | 0.75 (0.66- 0.85) |
| Rudolph[h] | 122 | 63(52) | - | 4 | 0.74 |
| PRE-DELIRIC[i] | | | mLR | 10 | 0.87(0.85- 0.89) |

Abbreviations: AUROC (Area under the receiver operating curve), P (Prospective Cohort), R (Retrospective cohort), MvLR (Multivariate Logistic Regression), mLR (Multiple Logistic Regression), LR (Logistic Regression), GLMM (General Linear Mixed Model), GBM (Gradient Boosting Machine), ANN (Artificial Neural Network), LSVM (Linear Support Vector Boosting Machine), RF (Random Forest), PRE-DELIRIC (The Prediction Model for Delirium).

[a] Temporal Split of cohort for development and validation
[b] Delirium incidence within individual cohorts was not reported.
[c] Random Split of cohort for development and validation.
[d] Recalibration of existing model
[e] Logistic Regression Machine Learning technique
[f] Moon et al performed three separate internal validations.
[g] This model applied machine learning techniques to EKGs. There are an unknown number of derived features.
[h] Validated model developed prior to 2013, which was included in van Meenen 2014.
[i] Model was validated (Azuma 2019, Green 2019, Lee 2017, Paton 2016, Wassenaar 2018) but not developed in any of the studies included in our review.

## 3.5 Validation of Models

Our review included six studies primarily focused on the validation of existing models [20, 35-39] (Table 3). These studies used logistic regression and calibration curves for their analysis with the exception of one [37] which did not include any statistical analysis. Additionally, many of the model development studies included internal validation of their models. Twelve of the 26 models were validated, either internally, through a split cohort, or externally using data from a separate institution [20, 22-25, 27, 35, 38, 40]. The remaining fourteen models either lacked any validation or merely utilized bootstrapping [26, 28-30, 32, 34, 36, 40]. Of the 11 internally validated models, cohorts were split either temporally or randomly. The reported AUROC of the externally validated models ranged from 0.68 to 0.83 [20, 38] with the majority of models exhibiting poorer performance during validation. Of the 15 models with an AUROC of 0.75 or greater in development, ten were validated in a split or separate cohort [24, 25, 27, 35, 38, 40] and only four maintained an average AUROC of 0.75 or higher during validation [24, 25, 35, 40]. Five studies externally validated the PRE-DELIRIC model [20, 35-38], two validated the E-PRE-DELIRIC [35, 38], three the recalibrated PRE-DELIRIC [35, 36, 39], and one the model proposed by Chen et al [23].

Table 3. Model Performance in Validation Studies

| Derivation Study | Validation Study | Validation Type | Sample Size (n) | Delirium Measure | Delirium identified by | Delirium n (%) | AUROC |
|---|---|---|---|---|---|---|---|
| PRE-DELIRIC[a] | Azuma | External | 70[b] | ICDS | Nurse, psychiatrist | 22 (20) | 0.83 |
| | Green | External | 445[b] | CAM-ICU | Unspecified | 160 (36) | 0.79 |
| | Lee | External | 600[c] | CAM-ICU; haloperidol | Nurse | 83 (14) | 0.75 |
| | Paton | External | 32[c] | CAM-ICU | Nurse | 15 (47) | - |
| | Wassenaar, 2018 | External | 2178[c] | CAM-ICU | Nurse | 467 (16) | 0.74 |
| E-PRE-DELIRIC[d] | Green | External | 445[b] | CAM-ICU | Unspecified | 160 (36) | 0.72 |
| | Wassenaar, 2018 | External | 2187[c] | CAM-ICU | Nurse | 467 (16) | 0.68 |
| rPRE-DELIRIC[e] | Green | External | 445[b] | CAM-ICU | Unspecified | 160 (36) | 0.79 |
| | Lee | External | 600[c] | CAM-ICU; haloperidol | Nurse | 83 (14) | 0.75 |
| | Linkaite | External | 38[c] | CAM-ICU | Researcher | 22 (58) | 0.713 |
| Chen, 2017 | Green | External | 445[b] | CAM-ICU | Unspecified | 160 (36) | 0.77 |

Abbreviations Used: AUROC (Area under the receiver operating curve), ICDS (Intensive Care Delirium Screening Checklist), CAM-ICU (Confusion Assessment Method for the Intensive Care Unit), PRE-DELIRIC (The Prediction Model for Delirium), E-PRE-DELIRIC (The Early Prediction Model for Delirium), rPRE-DELIRIC (The Recalibrated Prediction Model for Delirium).
[a] Model was developed prior to 2013.
[b] Retrospective Cohort
[c] Prospective Cohort
[d] Developed by Wassenaar et al, 2015.
[e] Recalibrated PRE-DELIRIC developed by van den Boogaard et al, 2014.

**3.6 Risk Prediction Performance**

Twenty models stratified patients into two to five risk groups using cut off values calculated with Youden's Index [22, 33]. Studies most commonly reported the sensitivity, specificity, positive predictive value, and negative predictive value associated with the determined threshold, though positive and negative likelihood ratios were also reported by some studies [22, 42]. While models varied as to whether they had a higher sensitivity (59% - 90.9%) [20, 26] or specificity (56.5% - 92.5%) [25, 34], the negative predictive value was generally much higher than the positive predictive value, indicating that it was more common for patients who would not develop delirium to be erroneously assigned to the high-risk group than for delirious patients to be erroneously assigned to the lower-risk group.

**4. Discussion**

**4.1 Summary of Findings**

This review identified 26 predictive models for delirium in an ICU setting, of which 18 were developed within the 20 included studies. Of the 26 models, five were externally validated by a separate study [23, 31, 33, 40, 42] and another ten were internally validated by a separate or split cohort [23-25, 27, 33, 40]. Five studies were retrospective and assessed delirium incidence through electronic health record data [20, 29, 30, 34, 35]. Studies had various methods of assessing delirium including CAM-ICU, CAM, NuDESC, DSM, and ISDC.

Models predicted delirium using various numbers of risk factors ranging from 1 to 588 [32, 34]. Thirteen of the 23 models, which reported the number of predictions in the final model, included six to eleven predictors [23-25, 27, 29, 31, 33, 36, 40]. Across the models reviewed, there was a significant amount of overlap between candidate and final predictors, including age, APACHE II, renal function, sedative use, benzodiazepine use, mechanical ventilation, and urgent admission (Figure 2). Despite the overlap between risk factors, there were differences in how predictors were defined and measured, making it difficult to directly compare the relative importance or weight of the predictor in each model. For example, mental status was assessed by MMSE score [40], history of dementia [23], use of Alzheimer's medication [26], and history of cognitive impairment [33].

Most studies utilized logistic regression in the development of their models, three studies used machine learning techniques [27, 28, 34], and two recalibrated existing models [31, 36]. The studies which utilized machine learning had larger cohorts and a greater number of candidate predictors. One of these models took the unique approach of applying machine learning techniques to analyze electrocardiograms to predict delirium [28]. Additionally, one developed a model aimed at predicting daily transitions between multiple states (normal, delirious, comatose, discharge, or death) rather than predicting the development of delirium at any time [26]. Twenty models and five validation studies reported AUROC for the development and validation of their model, with values ranging from 0.62 to 0.94 [25, 29]. Models tended to exhibit poorer discrimination in validation than in development, and calibration was inconsistently reported.

**4.2 Application to Practice**

Delirium has classically been described as a transient, waxing and waning condition. Interest in creating delirium prediction models emerges from clinicians' difficulties in recognizing the signs and risk factors of this multifactorial and dynamic condition that can evolve on an hourly basis. Patients admitted to the ICU already have critical medical conditions, the diagnosis, treatment, and monitoring of which demand clinicians' time and attention. Given these time constraints, uncertainty, and changing conditions, delirium is often unrecognized, or is recognized in a delayed fashion. The studies reviewed generally seemed to voice an appreciation for the challenge of predicting delirium amidst the high demands of the critical care setting. Thus, considerations such as the ease-of-use of the developed models were often noted. Wassenaar et al., for example, noted that ICU physicians rated the user convenience of E-PRE-DELIRIC superior to PRE-DELIRIC despite the latter having superior performance in predicting delirium [38]. However, the trade-off between the ease of implementing delirium prediction models in clinical practice and the actual predictive power or clinical utility of these models need not persist, given recent advances in automation and machine learning. Moon et al. created and implemented a delirium prediction algorithm in an electronic medical record system, which updated every day at midnight, making the system readily accessible to healthcare providers [27]. Unfortunately, the algorithm's low positive predictive value (0.52) may lead to alarm fatigue. This model, along with Marra et al.'s model, avoids the temporally static prediction paradigm that is a limiting factor of most ICU delirium prediction models [26]. The most frequently studied and cited models rely on a set of factors collected at a single time-point to predict whether delirium will occur at any point during the remainder of the ICU admission.

Such models are unable to account for the dynamic condition of patients and delirium itself, each of which can change on an hourly basis.

### 4.3 Strengths and Limitations

We conducted a thorough review of PubMed, Embase, Cochrane Central, Web of Science, and CINAHL databases to ascertain a comprehensive picture of what is available in the literature on delirium predictive models in the ICU. This review builds on the work of van Meenen et al. [40] by including articles published from 2014 through April 25th, 2019. This is important because machine learning approaches have emerged and evolved during these recent years. Other systematic reviews have exclusively studied older adults [43, 44], excluded validation studies [45], or were restricted to cardiac surgery patients [36]. The studies included in our review were summarized in detail in terms of candidate predictors, final model predictors, and risk of bias [21]. Limitations of our review include: limiting the inclusion criteria to studies published within the past five years, including only ICU delirium prediction models, and excluding non-English studies. Studies that developed and validated delirium prediction models that were not limited to the context of the ICU, especially in surgical patients, have used innovative methods such as electroencephalography [46, 47] and near field infrared spectroscopy [46] to predict delirium, which may provide valuable information when predicting ICU delirium for postoperative patients.

### 5. Conclusion

Many ICU delirium prediction models have been developed and validated within the last five years. Most of these models were developed with similar statistical methods and use common predictive factors, though inconsistencies in how these factors were assessed and used obviates a consensus, as does the risk of bias. External validation efforts have primarily focused on a few select models, especially PRE-DELIRIC and E-PRE-DELIRIC, making external validation of competing models an area where further research is needed. Most delirium prediction models use a single snapshot in time, usually within 24 hours of admission, and do not account for fluctuations in patients' conditions during ICU admission. This is inconsistent with critical illness and delirium pathophysiology. Further research is needed to create clinically relevant dynamic delirium prediction models, which can not only adapt over time, but deliver pragmatic and actionable predictions to clinicians.

16**References**

1. Setters B, Solberg LM, (2017) Delirium. Primary care 44: 541-559

2. Maldonado JR, (2017) Acute Brain Failure: Pathophysiology, Diagnosis, Management, and Sequelae of Delirium. Critical care clinics 33: 461-519

3. Kanova M, Sklienka P, Roman K, Burda M, Janoutova J, (2017) Incidence and risk factors for delirium development in ICU patients - a prospective observational study. Biomedical papers of the Medical Faculty of the University Palacky, Olomouc, Czechoslovakia 161: 187-196

4. Foroughan M, Delbari A, Said SE, AkbariKamrani AA, Rashedi V, Zandi T, (2016) Risk factors and clinical aspects of delirium in elderly hospitalized patients in Iran. Aging clinical and experimental research 28: 313-319

5. Mori S, Takeda JR, Carrara FS, Cohrs CR, Zanei SS, Whitaker IY, (2016) Incidence and factors related to delirium in an intensive care unit. Revista da Escola de Enfermagem da U S P 50: 587-593

6. Vasilevskis EE, Han JH, Hughes CG, Ely EW, (2012) Epidemiology and risk factors for delirium across hospital settings. Best practice & research Clinical anaesthesiology 26: 277-287

7. Pandharipande PP, Pun BT, Herr DL, Maze M, Girard TD, Miller RR, Shintani AK, Thompson JL, Jackson JC, Deppen SA, Stiles RA, Dittus RS, Bernard GR, Ely EW, (2007) Effect of sedation with dexmedetomidine vs lorazepam on acute brain dysfunction in mechanically ventilated patients: the MENDS randomized controlled trial. Jama 298: 2644-2653

8. Ouimet S, Kavanagh BP, Gottfried SB, Skrobik Y, (2007) Incidence, risk factors and consequences of ICU delirium. Intensive care medicine 33: 66-73

9. Van Rompaey B, Elseviers MM, Schuurmans MJ, Shortridge-Baggett LM, Truijen S, Bossaert L, (2009) Risk factors for delirium in intensive care patients: a prospective cohort study. Critical care (London, England) 13: R77

10. van de Pol I, van Iterson M, Maaskant J, (2017) Effect of nocturnal sound reduction on the incidence of delirium in intensive care unit patients: An interrupted time series analysis. Intensive & critical care nursing 41: 18-25

11. Slooter AJ, Van De Leur RR, Zaal IJ, (2017) Delirium in critically ill patients. Handbook of clinical neurology 141: 449-466
16

**Supplement A: Search Terms**
**PubMed**

(delirium[tiab] OR "Delirium"[Mesh] OR "ICU syndrome"[tiab] OR "Acute Brain Dysfunction"[tiab] OR "acute confusion"[tiab] OR CAM[tiab] OR "Confusion assessment method"[tiab] OR psychosis[tiab] OR "mental impairment"[tiab])

AND

(ICU OR "intensive care*" OR "Intensive Care Units"[Mesh] OR "critical care*" OR "Critical Care"[Mesh] OR "critically ill" OR CCU)

AND

("predict*"[tiab] OR "model*"[tiab] OR "Models, Statistical"[Mesh] OR "Neural Networks (Computer)"[Mesh] OR "risk model*"[tiab] OR "Statistical Model"[tiab] OR "Probabilistic Model"[tiab] OR "linear mod*"[tiab] OR "logistic*"[tiab] OR "risk as*"[tiab] OR "Logistic Models"[Mesh] OR "Risk assessment"[Mesh])

AND

English[lang]

AND

"loattrfull text"[sb]

AND

"2009/04/25"[PDat] : "2019/04/25"[PDat]

**Embase**

('intensive care unit'/exp OR 'gicu' OR 'gicus' OR 'icu`s' OR 'close attention unit' OR 'combined medical and surgical icu' OR 'combined surgical and medical icu' OR 'critical care unit' OR 'general icu' OR 'intensive care department' OR 'intensive care unit' OR 'intensive care units' OR 'intensive therapy unit' OR 'intensive treatment unit' OR 'medical-surgery icu' OR 'medical/surgical icu' OR 'medical/surgical icus' OR 'medico-surgical icu' OR 'mixed medical and surgical icu' OR 'mixed surgical and medical icu' OR 'respiratory care unit' OR 'respiratory care units' OR 'special care unit' OR 'surgery/medical icu' OR 'surgical-medical icus' OR 'surgical/medical icu' OR 'unit, intensive care' OR 'intensive care'/exp OR 'care, intensive' OR 'critical care' OR 'intensive care' OR 'intensive therapy' OR 'therapy, intensive' OR 'critically ill patient'/exp OR 'critically ill' OR 'critically ill patient')

AND

('delirium'/exp OR 'acute delirium' OR 'chronic delirium' OR 'delier' OR 'delire' OR 'deliria' OR 'delirious state' OR 'delirious syndrome' OR 'delirium' OR 'delirium acutum' OR 'intensive care psychosis'/exp OR 'icu psychosis' OR 'intensive care delirium' OR 'intensive care dementia' OR 'intensive care psychosis' OR 'acute confusion'/exp OR 'acute confusion' OR 'icu syndrome' OR 'acute brain dysfunction' OR 'confusion assessment method'/exp OR 'mental impairment')

AND

('prediction and forecasting'/exp OR 'prediction and forecasting' OR 'prediction'/exp OR 'prediction' OR 'model'/exp OR 'model' OR 'modeling' OR 'modelling' OR 'models, nursing' OR 'neural network'/exp OR 'statistical model'/exp OR 'likelihood functions' OR 'linear model' OR 'linear models' OR 'logistic models' OR 'models, econometric' OR 'models, statistical' OR 'statistical model' OR 'probabilistic neural network'/exp OR 'risk assessment'/exp OR 'assessment, safety' OR 'risk adjustment' OR 'risk analysis' OR 'risk assessment' OR 'risk evaluation' OR 'safety assessment' OR 'risk'/exp OR 'risk' OR 'risk hypothesis' OR 'prognostic model' or 'prognosis')

AND

(2009:py OR 2010:py OR 2011:py OR 2012:py OR 2013:py OR 2014:py OR 2015:py OR 2016:py OR 2017:py OR 2018:py OR 2019:py)

**Cochrane Central**
#1 ((delirium OR 'ICU syndrome' OR 'Acute Brain Dysfunction' OR 'acute confusion' OR CAM OR 'Confusion assessment method' OR psychosis OR 'mental impairment')):ti,ab,kw (Word variations have been searched)
#2 ((ICU OR 'intensive care' OR 'critical care' OR 'critically ill' OR CCU)):ti,ab,kw (Word variations have been searched)
#3 ((predict OR model OR "risk model" OR "Statistical Model" OR "Probabilistic Model" OR "linear mod" OR logistic OR "risk ass" OR "prognostic mod*")):ti,ab,kw (Word variations have been searched)
#4 MeSH descriptor: [Delirium] explode all trees
#5 MeSH descriptor: [Critical Illness] explode all trees
#6 MeSH descriptor: [Critical Care] explode all trees
#7 MeSH descriptor: [Models, Biological] explode all trees
#8 MeSH descriptor: [Computer Simulation] explode all trees
#9 MeSH descriptor: [Models, Theoretical] explode all trees
#10 MeSH descriptor: [Bayes Theorem] explode all trees
#11 MeSH descriptor: [Decision Support Systems, Clinical] explode all trees
#12 MeSH descriptor: [Decision Support Techniques] explode all trees
#13 MeSH descriptor: [Neural Networks (Computer)] explode all trees
#14 (#4 OR #1) AND (#2 OR #5 OR #6) AND (#3 OR #7 OR 8 OR #9 OR #10 OR #11 OR #12 OR #13) with Cochrane Library publication date Between Jan 2009 and Dec 2019

**Web of Science**
TS=(delirium* OR ICU syndrome OR Acute Brain Dysfunction OR acute confusion OR CAM OR Confusion assessment method OR psychosis OR mental impairment)
AND
TS=(ICU OR intensive care* OR critical care* OR critically ill OR CCU)
AND
TS=(predict* OR model* OR risk model* OR Statistical Model OR Probabilistic Model OR linear mod* OR logistic* OR risk ass* OR prognostic)
AND
(LA=(English))
AND
2009-2019[a]

[a]. Selection made from web interface

**CINAHL**

((TI "delirium*") OR (AB "delirium*") OR (MH "Delirium") OR (MH "ICU Psychosis") OR "ICU syndrome" OR "Acute Brain Dysfunction" OR "acute confusion" OR CAM OR "Confusion assessment method" OR psychosis OR "mental impairment")
AND
(ICU OR "intensive care*" OR "critical care*" OR "critically ill" OR CCU)
AND
("predict*" OR "model*" OR (MH "Models, Biological") OR (MH "Models, Theoretical") OR (MH "Computer Simulation") OR (MH "Cox Proportional Hazards Model") OR (MH "Risk Assessment") OR "risk model*" OR "Statistical Model" OR "Probabilistic Model" OR "linear mod*" OR "logistic*" OR "risk as*")
AND
English
AND
10 years

Supplementary Table 1. Model Predictors

| Study | Number of Predictors | Demographics | Clinical Scores | Labs and Other Measures | Medications | Clinical Condition and Illness Severity | Factors Related to Clinical Context | Patient Comorbidities | Physical Exam | Surgical Factors |
|---|---|---|---|---|---|---|---|---|---|---|
| Chaiwat, 2019 | 5 | Age | SOFA IQCODE | -- | Benzodiazepines | Mechanical ventilation | -- | -- | -- | -- |
| Chen, 2017 | 11 | Age | APACHE II | -- | Dexmedetomindine HCl | Metabolic acidosis Coma Mechanical ventilation Multiple trauma | Emergency operation | History of dementia History of delirium History of hypertension | -- | -- |
| Fan, 2019 | 7 | -- | APACHE II | -- | Analgesics Sedatives | Sleep disturbance | -- | History of chronic disease | Sensory deficits Restraint Indwelling catheter | -- |
| Kim, 2016 | 9 | Age | -- | Preoperative CRP | -- | -- | ICU admission | Heavy alcoholism History of delirium | Low physical activity Hearing impairments | Emergency Surgery Open surgery |
| Lee, 2017 | 6 | Age | -- | -- | Preoperative beta-blockers Preoperative statins | -- | -- | History of CVA/TIA Preoperative depression | -- | Surgery type |
| Marra, 2018 | 14 | ICU day number Age | SOFA APACHE II | -- | Opiates Propofol Antipsychotic agents Benzodiazepines | Sepsis Mechanical ventilation | ICU type | Medication use for Alzheimer's | Current brain function | -- |
| Moon, 2018 | 11 | Age Education | LOC score | BUN Psychopharmacological medication | -- | Infection | Medical department | -- | Pulse Activity level Number of catheters Restraints | -- |
| Oh, 2018 | 1[a] | -- | -- | ECG | -- | -- | -- | -- | -- | -- |
| Sakaguchi, 2018 | 6 | -- | -- | IVC diameter, higher TRPG Cr BUN CRP | | -- | -- | Previous incidence of OCI | -- | -- |

| Study | Number of Predictors | Demographics | Clinical Scores | Labs and Other Measures | Medications | Clinical Condition and Illness Severity | Factors Related to Clinical Context | Patient Comorbidities | Physical Exam | Surgical Factors |
|---|---|---|---|---|---|---|---|---|---|---|
| Stukenberg, 2016 | 3 | -- | PARS RASS NuDESC | -- | -- | -- | -- | -- | -- | -- |
| van den Boogaard, 2014 | 10 | Age | APACHE II | Urea concentration | Morphine Sedatives | Metabolic acidosis Infection Coma | Admission category Urgent admission | -- | -- | -- |
| Wang, 2018 | 1 | -- | STOP BANG | -- | -- | -- | -- | -- | -- | -- |
| Wassenaar, 2015 | 9 | Age | -- | BUN | Corticosteroids | Respiratory failure | Admission category Urgent Admission | History of cognitive impairment History of alcohol abuse | Mean arterial bp | -- |
| Wong, 2018 | | | | | | | | | | |
| Logistic regression[b] | 114 | | | | | | | | | |
| Gradient Boosting Machine[b] | 345 | | | | | | | | | |
| Artificial Neural Network[b] | Unspecified | | | | | | | | | |
| LSVM[b] | Unspecified | | | | | | | | | |
| Random Forest[b] | 588 | | | | | | | | | |
| Bohner, 2003[c] | 9 | Age | MMSE Hamilton Depression Rating Scale | -- | -- | -- | -- | History of subaortic occlusive disease History of major amputation History of hypercholesterolemia | Height | Colloid infusion Minimal potassium |

| Study | Number of Predictors | Demographics | Clinical Scores | Labs and Other Measures | Medications | Clinical Condition and Illness Severity | Factors Related to Clinical Context | Patient Comorbidities | Physical Exam | Surgical Factors |
|---|---|---|---|---|---|---|---|---|---|---|
| Katznelson, 2009[c] | 6 | Age | -- | -- | Preoperative beta-blockers Preoperative statins | -- | -- | History of CVA/TIA Preoperative depression | -- | Surgery type |
| Katznelson, 2009[c] | 7 | Age | -- | Preoperative Cr | Preoperative statins RBC transfusion | -- | -- | Preoperative depression | -- | Surgery type Intra-aortic balloon pump support |
| Marcantonio, 1994[c] | 6 | Age | Cognitive Status score RASS | Electrolyte disturbance | -- | -- | -- | Alcohol Abuse | -- | Surgery Type |
| Kalisvaart-Inouye, 2006[c] | 4 | -- | MMSE APACHE II | BUN/Cr ratio | -- | -- | -- | -- | Vision score | -- |
| Koster, 2012[c] | 2 | -- | Euroscore | Electrolyte disturbance | -- | -- | -- | -- | -- | -- |
| Rudolph, 2009[c] | 4 | -- | -- | MMSE Geriatric depression scale | Serum albumin | -- | -- | History of CVA/TIA | -- | -- |
| van den Boogaard, 2012[d] | 10 | Age | APACHE II | Urea concentration | Morphine sedative | Metabolic acidosis Infection Coma | -- | Admission category Urgent admission | -- | -- |

Abbreviations used: SOFA (Sequential Organ Failure Assessment), IQCODE (Informant Questionnaire of Cognitive Decline in the Elderly), APACHE II (Acute Physiology, Age, and Chronic Health Evaluation), CRP (C-Reactive Protein), ICU (Intensive Care Unit), CVA/TIA (Cerebrovascular Accident/ Transient Ischemic Attack), LOC (Level Of Consciousness), BUN (Blood Urea Nitrogen), ECG (electrocardiogram), IVC (Interior Vena Cava), TRPG (Tricuspid Regurgitation Peak Gradient), Cr (Creatinine), PARS (Post-Anesthetic Recovery Score), NuDESC (Nursing Delirium Screening Scale), STOP-BANG (Snoring, Tiredness, Observed apnea, blood Pressure, Body mass index, Age, Neck circumference, and Gender), bp (blood pressure), MMSE (Mini Mental State Evaluation), RBC (Red Blood Cell)

[a] Unknown number of derived features in model
[b] Multiple models developed in study by Wong et al.
[c] Studies included in literature review by van Meenen et al.
[d] This model was not developed in any of our included studies but was validated in Azuma et al, Green et al, Lee et al, Paton et al, and Wassenaar et al (2018).

Supplementary Table 2. Prevalence of Predictors Considered in Five or More Models

| Study | APACHE II | Markers of Renal Function | Anti-psychotics | Sedatives | Benzodiazepines | Mechanical Ventilation | Urgent Admission / Surgery | Age |
|---|---|---|---|---|---|---|---|---|
| Chaiwat | * | * | * | * | + | + | ‡ | + |
| Chen | + | ‡ | ‡ | ‡ | ‡ | + | + | + |
| Fan | + | ‡ | ‡ | + | ‡ | * | ‡ | * |
| Kim | ‡ | * | * | * | * | ‡ | * | + |
| Marra | + | ‡ | + | ‡ | + | + | ‡ | + |
| Sakaguchi | ‡ | + | ‡ | ‡ | * | + | ‡ | * |
| van den Boogaard | + | + | ‡ | + | ‡ | ‡ | + | + |
| Wassenaar 2015 | ‡ | + | * | ‡ | ‡ | ‡ | + | + |
| Wong | | | | | | | | |
|   Linear Regression | ‡ | # | # | # | # | # | # | # |
|   Gradient Boosting Machine | ‡ | # | # | # | # | # | # | # |
|   Artificial Neural Network | ‡ | # | # | # | # | # | # | # |
|   Linear Support Vector Machine | ‡ | # | # | # | # | # | # | # |
|   Random Forest | ‡ | # | # | # | # | # | # | # |

\+ predictor included in model

\* predictor considered but not used in final model

\# predictor considered but not specified whether it was used in model

‡ predictor not considered in model

Supplementary Table 3. Model performance in development studies

| Study (cohort type) | Development/ Validation | Delirium Measure (identified by) | Delirium n (%) Development/ Validation | Modelling Approach | Number of Predictors (Events per Variable)[a] | AUROC (95% CI) | Accuracy | Positive/ Negative Predictive Value | Predictors (weight) |
|---|---|---|---|---|---|---|---|---|---|
| Chaiwat (P) | 250/ - | CAM-ICU (nurse) | 61(24)/ NA | MvLR | 5 (2.9) | 0.84 (0.796- 0.897) | 0.92 | 55.0/ 90.0 | Age (0.057) <br> SOFA score (0.230) <br> Benzodiazepines (0.813) <br> Diabetes (1.109) <br> MV (1.178) <br> IQCODE score (1.219) |
| Chen (P) | 310/ 310[b] | CAM-ICU (nurse, doctor) | 160(26)[c] | MvLR | 11 (14.5) | - | - | - | Age (0.001) <br> APACHE II (0.015) <br> MC (0.801) <br> Emergent operation (0.358) <br> Coma (0.004) <br> Multiple trauma (0.148) <br> Metabolic acidosis (0.358) <br> History of hypertension (0.117) <br> History of delirium (1.377) <br> History of dementia (0.318) <br> Dexmedetomidine hydrochloride (-0.57) |
| Fan (P) | 336/ 224[d] | CAM-ICU (researcher) | 68(20)/ 46(20) | mLR | 7 (2.6) | 0.888 (0.845- 0.932) | - | - | Hearing deficit (9) <br> Indwelling catheter (8) <br> Infection (4) <br> Sedatives (3) <br> Sleep disturbance (3) <br> APACHE II (3) <br> History of chronic diseases (3) |
| Kim (P) | 561/ 553[b] | CAM (nurse) | 112(20)/ 99(18) | mLR | 9 (2.3) | 0.911 (0.88- 0.94) | 0.904 | 70.2/ 95.7 | Age <br> Low physical activity (2) <br> Alcoholism (1) <br> Hearing impairment (1) <br> History of delirium (2) <br> Emergent surgery (1) <br> Open surgery (2) <br> ICU admission (3) <br> CRP > 10 mg.dL (1) |

| Study (cohort type) | Development/ Validation | Delirium Measure (identified by) | Delirium n (%) Development/ Validation | Modelling Approach | Number of Predictors (Events per Variable)[a] | AUROC (95% CI) | Accuracy | Positive/ Negative Predictive Value | Predictors (weight) |
|---|---|---|---|---|---|---|---|---|---|
| Lee (P) | 600/ - | CAM-ICU; haloperidol use (nurse) | 83(14)/ - | LR[e] | 7 (11.8) | 0.62 | - | - | RBC transfusion<br>Intra-aortic balloon pump support<br>Preoperative depression<br>Preoperative creatinine<br>Age<br>Surgery type<br>Preoperative statins |
| Marra (P) | 810/ - | CAM-ICU: RASS (researcher) | 606 (75)/ - | mLR | 14 (40.4) | - | 0.734 (0.729-0.738) | 54.8/ 82.3 | ICU day<br>Age<br>SOFA<br>APACHE II<br>Opiates<br>Propofol<br>Antipsychotics<br>Benzodiazepines<br>Sepsis<br>MV<br>ICU type<br>Alzheimer's medication<br>Brain function |
| Moon (P) | 3284/ 325[b] | CAM-ICU (2 researchers) | 688(21)/ 48 (15) | LR[f] | 11 (0.16) | 0.89 | 0.79 | 49/ 96 | Age (0.04)<br>Education in years (-0.07)<br>LOC score (1.00)<br>Pulse (0.02)<br>Low Activity (1.02)<br>BUN (0.01)<br>Infection (0.52)<br>Catheters (0.20)<br>Restraint (0.46)<br>Psychopharmacology drugs (0.48) |
| Oh (P) | 94/ - | DSM 5; CAM-ICU (psychiatrist, nurse) | 39 (41)/ - | LSVM | -[g] (-)[h] | - | 0.75 (0.72-0.78) | - | NA |

| Study (cohort type) | Development/ Validation | Delirium Measure (identified by) | Delirium n (%) Development / Validation | Modelling Approach | Number of Predictors (Events per Variable)[a] | AUROC (95% CI) | Accuracy | Positive/ Negative Predictive Value | Predictors (weight) |
|---|---|---|---|---|---|---|---|---|---|
| Sakaguchi (R) | 120/ - | ICDSC (nurse) | 38(32) / - | MvLR | 6(1.3) | 0.885 | 0.885 | - | IVC diameter, higher TRPG<br>Cr<br>BUN<br>CRP<br>Previous incidence of OCI |
| Stukenberg (P) | 996/ - | CAM; CAM-ICU (researcher)[i] | 161(16)/ - | LR | 3 (53.7) | - | - | - | PARS, RASS, nuDESC |
| van den Boogaard (P) | 1824/ - | CAM-ICU (nurse) | 363(20) / - | LR[e] | 10 (36.3) | 0.76 (0.74-0.79) | - | - | Age (0.0183)<br>APACHE II (0.0272)<br>Coma (drug induced [0.2578], misc. [1.0721], combination [1.3361])<br>Admission category (medical[0.1446], trauma [0.5316], neuro [0.6516])<br>Infection (0.4965)<br>Metabolic acidosis (0.1378)<br>Morphine (0.01-7.1 mg [0.1926], 7.2-18.6 mg [0.0625], >18.6 mg [0.2414])<br>Sedatives (0.6581)<br>Urea concentration (0.0141)<br>Urgent admission (0.1891) |
| Wang (R) | 1692/ NA | CAM-ICU; RASS (unspecified) | 32(25) delirium 92(72) delirium and coma | LR | 1 | NA | - | - | STOP-BANG |
| Wassenaar | 1962/ 952[d] | CAM-ICU; haloperidol use (nurse) | 480(25)/ 208 (22) | mLR | 9 (25.9) | 0.76 (0.73- 0.78) | - | - | Age (0.025)<br>History of cognitive impairment (0.878)<br>History of alcohol abuse (0.505)<br>Admission category (trauma [ 1.218], medical [0.370], neuro [ 0.504])<br>Urgent admission (0.612)<br>MAP (-0.006)<br>Corticosteroids (0.283)<br>Respiratory failure (0.982) |

| Study (cohort type) | Development/ Validation | Delirium Measure (identified by) | Delirium n (%) Development/ Validation | Modelling Approach | Number of Predictors (Events per Variable)[a] | AUROC (95% CI) | Accuracy | Positive/ Negative Predictive Value | Predictors (weight) |
|---|---|---|---|---|---|---|---|---|---|
| Wong (R) | 1822/ - | CAM-ICU; nuDESC (ICU nurse) | 848(5)/ - | - | - (1.1) | - | - | | BUN (0.018) |
| | | | | LR [j, f] | 114 | Age<65: 0.885; Age>64: 0.799 | | 9.7/ 99.2 | NA |
| | | | | GBM [j] | 345 | Age<65: 0.856; Age>64: 0.804 | | 9.4/ 99.1 | NA |
| | | | | ANN [j] | NA | Age<65: 0.712; Age>64: 0.765 | | - | NA |
| | | | | LSVM [j] | NA | Age<65: 0.712; Age>64: 0.764 | | - | NA |
| | | | | RF [j] | 588 | Age<65: 0.849; Age>64: 0.807 | | 9.3/ 99.1 | NA |
| Bohner[k] | 153 | - | 60(44) | - | 9 | - | - | - | No history of supra-aortic occlusive disease<br>History of major amputation<br>No history of hypercholesterolemia<br>Age<br>Height<br>Hamilton Depression Scale<br>MMSE<br>Intraoperative colloids<br>Intraoperative minimum potassium |
| Katznelson[k] | 582 | - | 128(22) | - | 6 | 0.75 | - | - | History of CVA/TIA<br>Preoperative depression<br>Age<br>Preoperative beta blockers<br>Preoperative statins<br>Surgery type |
| Katznelson[k] | 1059 | - | 122(11) | - | 7 | 0.77 | - | - | RBC transfusion<br>Intra-aortic balloon pump support<br>Preoperative depression<br>Preoperative creatinine<br>Age<br>Surgery type<br>Preoperative statins |
| Marcantonio[k] | 1341 | - | 117(9) | - | 6 | 0.81 | - | - | Age<br>Alcohol abuse<br>Cognitive status score<br>RASS |

| Study (cohort type) | Development/ Validation | Delirium Measure (identified by) | Delirium n (%) Development / Validation | Modelling Approach | Number of Predictors (Events per Variable)[a] | AUROC (95% CI) | Accuracy | Positive/ Negative Predictive Value | Predictors (weight) |
|---|---|---|---|---|---|---|---|---|---|
| Kalisvaart-Inouye[k] | 603 | - | 74(12) | - | 4 | 0.74 | - | - | Electrolyte disturbance<br>Surgery type<br>MMSE<br>APACHE II |
| Koster[k] | 300 | - | 52(17) | - | 2 | 0.75 | - | - | Vision score<br>BUN/Cr ratio<br>Euroscore |
| Rudolph[k] | 122 | - | 63(52) | - | 4 | 0.74 | - | - | Electrolyte disturbance<br>MMSE<br>History of CVA/TIA<br>Geriatric depression scale<br>Serum Albumin |
| PRE-DELIRIC[i] | | CAM-ICU | | mLR | 10 | 0.87 (0.85- 0.89) | - | - | Age (0.0387)<br>APACHE II (0.0575)<br>Coma (drug induced [0.5458], misc. [2.2695], combination [2.8283])<br>Admission category (medical[0.3061], trauma [1.1253], neuro [1.3793])<br>Infection (1.0509)<br>Metabolic acidosis (0.2918)<br>Morphine (0.01-7.1 mg [0.4078], 7.2-18.6 mg [0.1323], >18.6 mg [0.5110)<br>Sedatives (1.3932)<br>Urea concentration (0.0298)<br>Urgent admission (0.4004) |

Abbreviations: AUROC (Area Under the Receiver Operating Curve), P (Prospective Cohort), R (Retrospective cohort), MvLR (Multivariate Logistic Regression), mLR (Multiple Logistic Regression), LR (Logistic Regression), ML (Machine Learning), GBM (Gradient Boosting Machine), ANN (Artificial Neural Network), LSVM (Linear Support Vector Boosting Machine), RF (Random Forest), CAM-ICU (Confusion Assessment Method for the Intensive Care Unit), CAM (Confusion Assessment Method), RASS (Richmond Agitation and Sedation Score), DSM 5 (Diagnostic and Statistical Manual of Mental Disorders), ICDSC (Intensive Care Delirium Screening Checklist), nuDESC (Nurses Delirium Screening Score), PRE-DELIRIC (The Prediction Model for Delirium), SOFA (Sequential Organ Failure Assessment), MV (Mechanical Ventilation), IQCODE (Informant Questionnaire of Cognitive Decline in the Elderly), APCACHE II (Acute Physiology, Age, and Chronic Health Evaluation), ICU (Intensive Care unit), CRP (C-Reactive Protein), RBC (Red Blood Cell), LOC (level of Consciousness), BUN (Blood Urea Nitrogen), IVC (Interior Vena Cava), Cr (Creatinine), PARS (Post-Anesthetic recovery Score), RASS (Richmond Agitation and Sedation Scale), STOP-BANG (Snoring, Tiredness, Observed apnea, blood Pressure, Body mass index, Age, Neck circumference, and Gender), MAP (Mean Arterial Pressure), MMSE (Mini Mental Status Evaluation), CVA/ TIA (Cerebrovascular Attack/ Transient Ischemic Attack)

[a] Events per variable is the ratio of cases of delirium to the number of candidate predictors. Events per variable is not reported for models found in the literature review by van Meenen et al for they did not report the number of candidate predictors considered in the development of their included models.

[b] Temporal split of cohort for development and validation

[c] This study reported only its delirium incidence overall and did not specify delirium incidence for individual cohorts.

[d] Random split of cohort for development and validation

[e] Logistic regression used to recalibrate an existing model

[f] Logistic Regression machine learning techniques

[g] Unknown number of features derived

[h] EPV could not be calculated because number of candidate predictors is not specified.

[i] Under supervision of a psychiatrist and delirium specialist

[j] Separate models developed by Wong, 2018. See row entitled "Wong" for sample, delirium measure, delirium cases, and EPV data.

[k] Model was not developed in one of the included studies but data regarding its development was included in van Meenen, 2014. Assessment of delirium is not included for these models due to it not being reported in van Meenen 2014.

[l] Model was validated (Azuma 2019, Green 2019, Lee 2017, Paton 2016, Wassenaar 2018) but not developed in any of the studies included in our review.